# Water and Ice Dielectric Spectra Scaling at 0 °C

V.G. Artemov, A.A. Volkov

*A. M. Prokhorov General Physics Institute, Russian Academy of Sciences, 119991 Moscow, Russia*

Dielectric spectra ($10^4$-$10^{11}$ Hz) of water and ice at 0 °C are considered in terms of proton conductivity and compared to each other. In this picture, the Debye relaxations, centered at $1/\tau_W \approx 20$ GHz (in water) and $1/\tau_I \approx 5$ kHz (in ice), are seen as manifestations of diffusion of separated charges in the form of $H_3O^+$ and $OH^-$ ions. The charge separation results from the self-dissociation of $H_2O$ molecules, and is accompanied by recombination in order to maintain the equilibrium concentration, $N_\pm$. The charge recombination is a diffusion-controlled process with characteristic lifetimes of $\tau_W$ and $\tau_I$, for water and ice respectively. The static permittivity, $\varepsilon(0)$, is solely determined by $N_\pm$. Both, $N_\pm$ and $\varepsilon(0)$, are roughly constant at the water-ice phase transition, and both increase, due to a slowing down of the diffusion rate, as the temperature is lowered. The transformation of the broadband dielectric spectra at 0 °C with the drastic change from $\tau_W$ to $\tau_I$ is mainly due to an abrupt (by 0.4 eV) change of the activation energy of the charge diffusion.

## INTRODUCTION

At 0 °C water and ice coexist in a thermodynamic equilibrium. This makes it possible to correctly compare their electric properties and search for the microscopic mechanisms responsible for these properties. Both substances are dielectrics with a wide electronic band gap, ~ 5 eV [1]. For simple dielectrics, they exhibit high proton conductivity, $\sigma_{dc} \sim 10^{-7}$ $\Omega^{-1}$ cm$^{-1}$ for water, and three orders of magnitude smaller for ice. The dielectric constants are also anomalously high, $\varepsilon(0) \approx 90$ for both substances at frequencies below $10^3$ Hz. The identity of the dielectric constants is puzzling, because, as it is believed, the microscopic mechanisms of $\varepsilon(0)$ in water and ice are microscopically different. The dielectric constant $\varepsilon(0)$ of water is accepted to be due to reorientations of the molecular $H_2O$ dipoles, while the $\varepsilon(0)$ of ice is better understood *via* inter-oxygen (O-O) proton hopping [2, 3].

The dielectric spectra of water and ice at 0 °C are presented in Fig. 1 in terms of the dielectric permittivity, $\varepsilon'(\omega)$ and $\varepsilon''(\omega)$, and conductivity, $\sigma(\omega)$. The graphs are constructed from data taken selectively from recognized sources [2, 4, 5]. The spectra of the top panel are formed by two loss bands, $\varepsilon''(\omega)$, accompanied by the two steps of $\varepsilon'(\omega)$. This spectral anomaly, the peak of $\varepsilon''(\omega)$ and the step of $\varepsilon'(\omega)$ at characteristic frequency $\nu = 1/2\pi\tau_D$, is a well-known property of water and ice electrodynamics, known as the Debye relaxation. Its manifestations in the dielectric spectra of water and ice, as is seen, are very similar: they are the same graphics, shifted along the frequency axis by six decades.

Surprisingly, the observed similarity is still not explained. Traditionally, the dielectric properties of water and ice are studied in comparison but separately [1, 2]. In modern computer simulations this tradition persists [6, 7]. The first attempts to unify the approaches appeared quite recently [8].

In the present paper, we consider the problem from the standpoint of the newly developed concept of proton transport for liquid water [9]. We show that the assumption of a random walk diffusion of protons without considering the $H_2O$ dipole orientations is sufficient to describe both the experimental dielectric spectra of water and ice, as well as their transformation at the freezing-melting point of 0 °C.

## REMARKS ON EXISTING EXPERIMENTAL DATA

In terms of conductivity $\sigma(\omega)$, the graphs of the top panel in Fig. 1 (the two peaks of the dielectric loss $\varepsilon''(\omega)$) are transformed into two spectral knees in the bottom panel (at ~ 10 GHz for water and ~ 3 kHz for ice). Both the water and ice spectra flatten at the high- and low-frequency limits and are transformed into dispersionless plateaus, $\sigma_\infty$ and $\sigma_0$. The ratio of $\sigma(\omega)$ for water and ice oscillates twice during the spectrum around unit. The dynamic conductivity of water at frequencies of $10^7$-$10^{10}$ Hz is several orders of magnitude higher than the conductivity of ice, while in the range of $10^4$-$10^6$ Hz, the conductivity of ice exceeds by ~ 50 times the conductivity of water.

A remarkable feature of the Debye relaxation is that it is perfectly represented both for water and ice by the simple analytic form $\varepsilon^*(\omega)=\varepsilon_\infty+\Delta\varepsilon_D/(1-i\omega\tau_D)$ or separately for the real and imaginary parts:

$$\varepsilon'(\omega) = \varepsilon_\infty + \frac{\Delta\varepsilon_D}{1+\omega^2\tau_D^2}, \quad \varepsilon''(\omega) = \omega\tau\frac{\Delta\varepsilon_D}{1+\omega^2\tau_D^2}, \quad (1)$$

where $\tau_D$ is the relaxation time and $\Delta\varepsilon_D = \varepsilon(0) - \varepsilon_\infty$ is the contribution of the dielectric relaxation to the static dielectric constant $\varepsilon(0)$ ($\varepsilon_\infty$ is the high frequency limit, equaling 5 and 3 for water and ice, respectively). The Debye relaxation looks in terms of $\sigma(\omega)$ form like:

$$\sigma(\omega) = \omega^2\tau_D\varepsilon_0 \frac{\Delta\varepsilon_D}{1+\omega^2\tau_D^2}. \quad (2)$$

At high frequencies, $\sigma(\omega)$ is frequency-independent, corresponding to the plateau:

$$\sigma_\infty(\omega) = \sigma(\omega \to \infty) = \frac{\varepsilon_0\Delta\varepsilon_D}{\tau_D}. \quad (3)$$

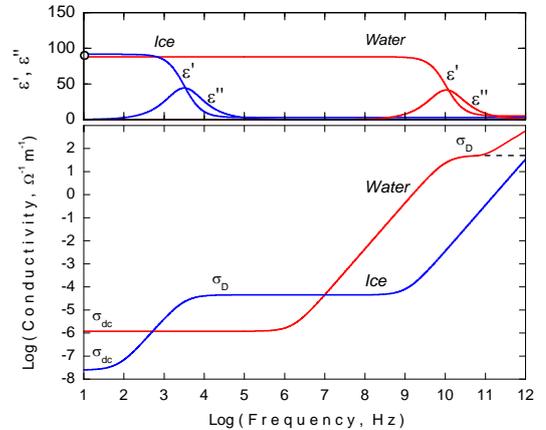

FIG.1: The dielectric spectra of water and ice at 0° C: the real and imaginary parts of permittivity, $\varepsilon'(\omega)$ and $\varepsilon''(\omega)$ (top panel), and the dynamical conductivity $\sigma(\omega)$ (bottom panel). The circle at the vertical axis indicates the static dielectric constant $\varepsilon(0) \approx 90$ (both for water and ice, within 10%).

The parameters of the model (1)-(3) are found in the literature to a great extent, since they have many times been repeatedly measured experimentally, both for water and ice. We are based upon data taken from [2, 4, 10]. The data for $\Delta\varepsilon_D$ and $\tau_D$ are taken as initial, while data for $\sigma_\infty$ are calculated in accordance with (3); their values at a temperature of 0 °C are given in Table 1. In addition, their temperature dependencies are included in our analysis. They are presented by the Arrhenius form, $A = B\,exp(\Delta E/kT)$, with the parameters $B$ and $\Delta E$, which are listed in Table 2. The activation energies $\Delta E$ are taken as the slopes of the Arrhenius straight lines (drawn in a range of 0 - 100 °C). The pre-exponential factors $B$ are fitted so that the calculated $A$ would be equal to their values in Table 1.

TABLE 1: The Debye relaxation parameters for water and ice according to [2, 4, 5, 10] and formula (3) at T ≈ 275 K: the dielectric contribution, $\Delta\varepsilon_D$; the lifetime of the separated charges, $\tau_D$; the limiting high-frequency conductivity, $\sigma_\infty$.

|  | $\Delta\varepsilon_D$ | $\tau_D$, s | $\sigma_D = \sigma_\infty$, $\Omega^{-1}m^{-1}$ |
|---|---|---|---|
| Water | 83 | $1.7\times10^{-11}$ | 43 |
| Ice | 93 | $2.2\times10^{-5}$ | $4\times10^{-5}$ |

TABLE 2: The Arrhenius coefficients of the temperature dependences $A(T)$ of the parameters, presented in Table 1: pre-exponential factor $B$ and activation energy $\Delta E$.

|  | $A(T)$ | $\Delta\varepsilon_D$ | $\tau_D$, s | $\sigma_D = \sigma_\infty$, $\Omega^{-1}m^{-1}$ |
|---|---|---|---|---|
| Water | $B$ | 14.0 | $1.5\times10^{-14}$ | $8.0\times10^3$ |
|  | $\Delta E$ | 0.042 | 0.165 | -0.123 |
| Ice | $B$ | 15.5 | $0.6\times10^{-15}$ | $2.5\times10^5$ |
|  | $\Delta E$ | 0.042 | 0.570 | -0.530 |

It should be noted that the phenomenological Debye model (1)-(3) is microscopically ambiguous [11]. This makes it possible to use different microscopic models to describe similar relaxation features in dielectric spectra. In particular, for water, the form (1) is commonly used to monitor the $\tau_D$ and $\Delta\varepsilon_D$ temperature changes, while for ice, the form (2), is used to describe the temperature behavior of $\tau_D$ and $\sigma_\infty$. A contradiction arises at 0 °C in that the use of (1) implies the high stability of a $H_2O$ molecule (when the dipole moments of intact $H_2O$ are considered) [1], while the use of (2), on the contrary, requires the instability of a $H_2O$ molecule (when proton hopping is considered) [12]. The compromise is that proton hops are equivalent at the end to $H_2O$ molecule reorientations [8].

**ARGUMENTS FOR OUR MODEL**

We introduced the conductivity for water, $\sigma(\omega)$, as a response of charges produced by the self-dissociation of $H_2O$ molecules [9]. In general, this mechanism is well known and thought to be responsible for the dc-conductivity in water. It is commonly assumed to be valid up to the frequencies of $10^7$ Hz. However, in [9] we extended its validity to much higher frequencies, up to ~ $10^{11}$ Hz, and showed that this makes it possible to describe both the Debye relaxation (~ $10^{10}$ Hz) and the dc-conductivity. The basic assumption is that the charges in the form of $H_3O^+$ and $OH^-$ ions of $N_\pm$ concentration randomly walk with the diffusion coefficient $D_\pm$. The excess or lack of a charge on an initially neutral $H_2O$ molecule is due materialistically to proton hopping. Thus in our model the conductivity $\sigma$, being proportional to $D_\pm$, is an indicator of proton motion.

The existence of separated charges implies their generation and recombination, which necessarily manifests itself in the electrodynamic response. Logically, the recombination of $H_3O^+$ (or $OH^-$) ion can occur with either its own or a foreign partner. The first meeting takes place through the time $\tau_D$ at a distance of $\ell_\pm$, while the second occurs through the time $\tau_L$ at a distance of $L_\pm$, where $L_\pm$ is the radius of the first coordination sphere of the system of separated charges. In the first case, recombination is rapid, while in the second, recombination is much longer. As a whole, the system of short- and long-lived charges is in thermodynamic equilibrium.

We described foreword scenarios through a set of diffusion formulas that connected the microscopic parameters $N_\pm$, $D_\pm$, $\ell_\pm$, $L_\pm$ and $\tau_L$, with the experimentally measured $\Delta\varepsilon_D$ and $\tau_D$. In particular:

$$N_\pm = K_N \Delta\varepsilon_D^3, \qquad (4)$$

$$l_\pm = K_l \frac{1}{\Delta\varepsilon_D}, \qquad (5)$$

$$D_\pm = K_D \times \frac{1}{\tau_D} \times \frac{1}{\Delta\varepsilon_D^2}, \qquad (6)$$

where $K$ subscripted by $N$, $l$, and $D$ indexes are temperature dependent coefficients. As is seen, $N_\pm$, $\ell_\pm$ and $D_\pm$ are simple functions of the input macroscopic parameters, $\tau_D$ and $\Delta\varepsilon_D$. The latter, in turn, *via* the equations (1) and (2) can be fitted to the experiment. This has been done in [9] and it has been demonstrated that the model fits comprehensively for the dielectric spectra of water (red graph in Fig. 1).

TABLE 3: The microscopic parameters for water and ice according to the formulas (5)-(10) at T ≈ 275 K: the charge concentration, $N_\pm$; the diffusion coefficient, $D_\pm$; the diffusion length, $l_\pm$.

|  | $N_\pm$, m$^{-3}$ | $D_\pm$, m$^2$/s | $l_\pm$, nm |
|---|---|---|---|
| Water | $6\times10^{26}$ | $5.2\times10^{-9}$ | 0.75 |
| Ice | $8\times10^{26}$ | $3.2\times10^{-15}$ | 0.65 |

TABLE 4: The Arrhenius coefficients of the temperature dependences of the parameters, presented in Table 3: pre-exponential factor $B$ and activation energy $\Delta E$.

|  | $A(T)$ | $N_\pm$, m$^{-3}$ | $D_\pm$, m$^2$/s | $l_\pm$, nm |
|---|---|---|---|---|
| Water | $B$ | $3.4\times10^{26}$ | $1.9\times10^{-5}$ | $1.4\times10^9$ |
|  | $\Delta E$ | 0.042 | -0.193 | -0.014 |
| Ice | $B$ | $4.5\times10^{26}$ | $3.6\times10^{-4}$ | $1.2\times10^9$ |
|  | $\Delta E$ | 0.042 | -0.598 | -0.014 |

In view of the fact that the Debye relaxation is inherent for water in both its liquid and solid phases, the $N_\pm$, $\ell_\pm$ and $D_\pm$ values (4)-(6) can be compared directly for water and ice at 0 °C (the parameters for ice are marked by a dash):

$$N'_\pm = \left(\frac{\Delta\varepsilon'_D}{\Delta\varepsilon_D}\right)^3 N_\pm, \qquad (7)$$

$$l'_\pm = l_\pm \frac{\Delta\varepsilon_D}{\Delta\varepsilon'_D}, \qquad (8)$$

$$D'_\pm = \left(\frac{\Delta\varepsilon_D}{\Delta\varepsilon'_D}\right)^2 \frac{\tau_D}{\tau'_D} D_\pm. \qquad (9)$$

The numerical data are presented in Tables 3 and 4. Let us note especially that at 0 °C $N'_\pm = 1.4 \times N_\pm$; $\ell'_\pm = 0.9 \times \ell_\pm$; $D'_\pm = 0.8 \times 10^{-6} \times D_\pm$.

## DISCUSSION

As is seen from (7) and (8), the two main microscopic parameters, the concentration $N_\pm$ of $H_3O^+$-$OH^-$ ion pairs and their size $\ell_\pm$, are related for water and ice *via* solely the ratio of the dielectric contributions $\Delta\varepsilon_D$ and $\Delta\varepsilon'_D$. The ratio $\Delta\varepsilon_D/\Delta\varepsilon'_D$ is of an order of the unit (within 10% [12]), thus showing that $N_\pm$ and $\ell_\pm$ do not change essentially at 0 °C. Given that we conclude that both water and ice contain dissociated $H_2O$ molecules ($H_3O^+$-$OH^-$ ionization defects) at the same and very high concentration, $N_\pm \sim$ 1%. According to (8), the diffusion recombination area $\ell_\pm$ is also equal in water and ice.

At the same time, the diffusion coefficients $D_\pm$ and $D'_\pm$ are drastically different (by a factor of $10^6$). It follows from (9) that the $D_\pm \to D'_\pm$ transfer with the coefficient $\Delta\varepsilon_D/\Delta\varepsilon'_D \sim 1$ results in the $10^6$ multiple transfer of $\tau_D$ ($\tau_D \to \tau'_D$). This simultaneous $D_\pm$-$\tau_D$ transformation converts the red spectra of water in Fig. 1 into the blue spectra of ice ($D_\pm$-$\tau_D$ spectra scaling). Scaling clearly reveals that the dielectric response both in water and ice is due to the same microscopic mechanism, namely (in our model), to Brownian diffusion of opposite charges.

It should be noted that the analytic Brown-Einstein relationship between $D_\pm$ and $\tau_D$ (the direct proportionality of $D_\pm$ to $1/\tau_D$) results from the $\sigma_\infty$ to $1/\tau_D$ proportionality (3) caused by the Debye relaxation. Thus, the Debye band occurrence in the dielectric spectra of water and ice with the same static permittivity $\varepsilon(0)$ implies the $D_\pm$-$\tau_D$ scaling and reveals Brownian diffusion as the most adequate microscopic mechanism.

The Arrhenius temperature dependencies of the input parameters $\Delta\varepsilon_D$ and $\tau_D$ result in Arrhenius temperature dependencies for the microscopic parameters (7)-(9) - Table 3. The first striking result is a nontrivial temperature dependence of the concentration of dissociated $H_2O$ molecules, $N_\pm$ (the concentration of $H_3O^+$-$OH^-$ pairs). This value weakly reacts to the phase transition at 0 °C and monotonously increases (not freezes!) with activation energy -0.042 eV during temperature decreasing. At 100 K, at which the dielectric constant exceeds by two times the room temperature value [13], the concentration of defects in our model reaches 10%. This finding is in agreement with accepted opinion concerning the highly defected structure of ice and the non-zero entropy of ice at 0 K [1, 12].

Figure 2 shows the Arrhenius temperature dependences of $1/\tau_D$ with activation energy $\Delta_W = 0.165$ eV for water and $\Delta_I = 0.570$ eV for ice [2, 12]. At point 273 °K, the graphs show $\tau'_D/\tau_D = 10^6$ which corresponds to the frequency shift of the Debye bands in the dielectric spectra of ice and water in Fig. 1. In view of the $D_\pm$-$\tau_D$ scaling and in agreement with (9), the energy $\Delta = \Delta_I - \Delta_W = 0.405$ eV is dominantly responsible for a jump of the diffusion coefficient $D_\pm$ (gives $10^7$ contribution in comparison with 10 of the pre-exponential factor). By neglecting with 10 one can say that the activation energy of diffusion is a main (near-sole) factor of dielectric spectra transformation at 0 °C. This reveals that the proton diffusion is held at the water-ice phase transition, the only distinction between water and ice being the height of the potential barriers that the charges must overcome at each hop.

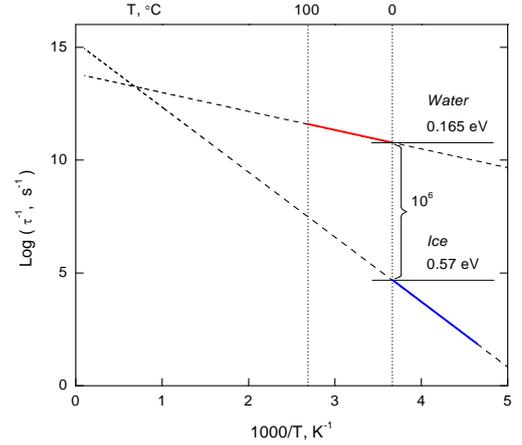

FIG. 2: The Arrhenius temperature dependences of the inverse Debye relaxation times of water and ice, $1/\tau_W$ and $1/\tau_I$ (digits with "eV" are the activation energies in electron volts).

The scheme described is illustrated in Fig. 3. Shown is an in-line stretched diffusion path of a charge starting from its birth to its recombination. The path is restricted by $n = (\ell/a)^2 \sim 9$ hops. The height of the barriers is $0.570/0.165 = 3.45$ times larger for ice. The first barrier includes the formation energy of a charge. For water, in particular, the first barrier height is $0.042 + 0.193 \approx 0.23$ eV (23 kJ / mol). It is noteworthy that the energy edition 0.042 eV to the first barrier multiplies its height by six (jump 1 in Fig. 3). About the same amount of jumps a charge makes to recombine (diffusion process 2). Thus, the birth and death of charges are consistent in our model. This meets the requirement of thermodynamic equilibrium and explains the above-noted specific thermal behavior of the concentration $N_\pm$ (the increase of $N_\pm$ with a temperature decrease). In fact, temperature lowing decelerates diffusion thus increasing the lifetime of the charges, as well as, correspondingly, their concentration $N_\pm$. According to (4), the growth of $N_\pm$ results in the growth of static permittivity $\varepsilon(0)$.

Let us note that our six-step recombination run for charges, which produces $\Delta\varepsilon_D$, comes from independent spectroscopic and thermal measurements. It is also consistent with the most probable six-element hydrogen-bonded loop in the computer simulation which fits calculations to experimental data concerning the $\varepsilon(0)$ of water and ice [8, 13].

The charge transfer in Fig. 3 is a set of consequent stages of the local redistribution of protons and electron clouds, therefore, it is, in essence, a diffusion-controlled chemical reaction. The difference between the total activation energies of the reactions in water and ice is $\Delta\Sigma = n \times \Delta \times N_\pm \approx 2 \times 10^{27}$ eV/m$^3$, which is close to the reference value of the latent heat of the water-ice phase transition, ~ $3 \times 10^2$ kJ/kg [1].

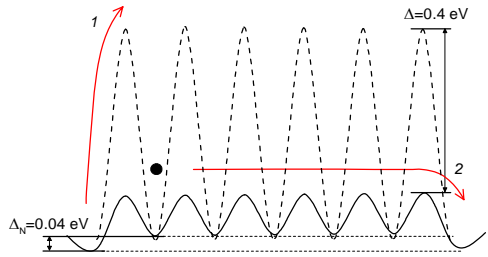

FIG. 3: The energy landscape of the charge diffusion in water (solid line) and ice (dashed line). The black circle is an excess proton in the $H_3O^+$ state (a charge), 1 - the charge activation process, 2 – the recombination process (diffusion), $\Delta_N$ is the charge activation energy, $\Delta$ is the n×$N_\pm$-th portion of the latent heat of fusion.

## CONCLUSION

Comparison of the dielectric spectra of water and ice at 0 °C in the framework of the model, represented by Eqs. (1)-(9), reveals the high degree of freedom of the proton motion, which appears to be entirely responsible for the low-frequency electrodynamics of water (at frequencies below $10^{11}$ Hz) and of ice (below $10^8$ Hz). Protons randomly walk by hopping over neutral $H_2O$ molecules to form short-lived $H_3O^+$-$OH^-$ ion pairs. The separated charges occur in water and ice in an unexpectedly high concentration, $N_\pm$, which accounts for the high value of the static dielectric constant, $\varepsilon(0)$. The equilibrium of the separated charges is dynamic. It is maintained by two competitive mechanisms – the self-dissociation of $H_2O$ molecules and recombination of the $H_3O^+$-$OH^-$ ion pairs. A decrease in temperature slows down the proton diffusion, thus shifting the equilibrium to increase the lifetime of the separated charges. This increases both, $N_\pm$ and $\varepsilon(0)$, while the overall temperature behavior of $N_\pm$ and $\varepsilon(0)$ is preserved at the water-ice phase transition.

According to the model, at 0 °C water and ice form very similar structures on scales within a dozen of $H_2O$ molecules, the only difference being the potential-barrier height. The total difference between the activation energies of the proton diffusion in water and ice is equal to the latent heat of the water-ice phase transition.

The authors thank A.V. Pronin for close collaboration on related topics.